\documentclass{article}

\usepackage[usenames,dvipsnames,svgnames,x11names]{xcolor}

\usepackage[usenames,dvipsnames,svgnames,x11names]{xcolor}

\newcommand{\hilite}[1]{{\textcolor{BlueViolet}{#1}}}

\usepackage{balance}

\usepackage[T1]{fontenc}
\usepackage[normalem]{ulem}

\usepackage[nocompress]{cite}

\usepackage{listings}

\lstdefinelanguage{XML}
{
basicstyle=\ttfamily\footnotesize,
  morestring=[b]",
  moredelim=[s][\bfseries\color{Maroon}]{<}{\ },
  moredelim=[s][\bfseries\color{Maroon}]{</}{>},
  moredelim=[l][\bfseries\color{Maroon}]{/>},
  moredelim=[l][\bfseries\color{Maroon}]{>},
  morecomment=[s]{<?}{?>},
  morecomment=[s]{<!--}{-->},
  commentstyle=\color{gray},
  stringstyle=\color{blue},
  identifierstyle=\color{red}
}
\usepackage{moreverb}

\usepackage[nounderscore]{syntax}

\usepackage[pdftex]{graphicx}
\graphicspath{{./figures/}}
\DeclareGraphicsExtensions{.pdf}

\usepackage[cmex10]{amsmath}
\usepackage{amssymb}
\usepackage{mathtools}
\usepackage{amsthm}
\usepackage{amsfonts}

\usepackage{subfig} 
\usepackage{makecell}

\usepackage{algorithmicx}
\usepackage{algpseudocode}
\usepackage[ruled]{algorithm}
\definecolor{light-gray}{gray}{0.75}
\algrenewcommand{\algorithmiccomment}[1]{\hskip3em{{\footnotesize \textcolor{light-gray}{$\blacktriangleright$}}} #1}
\usepackage{multirow} 
\usepackage{rotating} 
\usepackage{booktabs} 
\usepackage{colortbl} 
\usepackage{tablefootnote} 

\usepackage[pdftex,colorlinks=true,urlcolor=blue,citecolor=blue]{hyperref}

\usepackage{xspace}

\usepackage{enumitem}

\usepackage{blindtext}

\usepackage{listings}
\usepackage{xcolor}

\definecolor{codegreen}{rgb}{0,0.6,0}
\definecolor{codegray}{rgb}{0.5,0.5,0.5}
\definecolor{codepurple}{rgb}{0.58,0,0.82}
\definecolor{backcolour}{rgb}{0.95,0.95,0.92}

\lstdefinestyle{mystyle}{
    backgroundcolor=\color{backcolour},   
    commentstyle=\color{codegreen},
    keywordstyle=\color{magenta},
    numberstyle=\tiny\color{codegray},
    stringstyle=\color{codepurple},
    basicstyle=\ttfamily\footnotesize,
    breakatwhitespace=false,         
    breaklines=true,                 
    captionpos=b,                    
    keepspaces=true,                 
    numbers=left,                    
    numbersep=5pt,                  
    showspaces=false,                
    showstringspaces=false,
    showtabs=false,                  
    tabsize=2
}

\begin{document}
\date{}


\title{
\textit{AeroDaaS:} Towards an Application Programming Framework for Drones-as-a-Service
}

\author{Suman Raj, Rajdeep Singh, Kautuk Astu and Yogesh Simmhan\\
Department of Computational and Data Sciences, \\Indian Institute of Science, Bangalore 560012 India\\
Email: \{sumanraj, rajdeepsingh, kautukastu, simmhan\}@iisc.ac.in
}

\maketitle

\thispagestyle{plain}
\pagestyle{plain}

\begin{abstract}
The increasing adoption of UAVs with advanced sensors and GPU-accelerated edge computing has enabled real-time AI-driven applications in fields such as precision agriculture, wildfire monitoring, and environmental conservation. However, integrating deep learning on UAVs remains challenging due to platform heterogeneity, real-time constraints, and the need for seamless cloud-edge coordination.
To address these challenges, we introduce \textit{AeroDaaS}, a service-oriented framework that abstracts UAV-based sensing complexities and provides a \textit{Drone-as-a-Service} (DaaS) model for intelligent decision-making. AeroDaaS offers modular service primitives for on-demand UAV sensing, navigation, and analytics as composable microservices, ensuring cross-platform compatibility and scalability across heterogeneous UAV and edge-cloud infrastructures. We implement and evaluate a preliminary version of AeroDaaS for two real-world DaaS applications. We require $\leq40$ lines of code for the applications and see minimal platform overhead of $\leq20$ ms per frame and $\leq0.5$ GB memory usage on Orin Nano. These early results are promising for AeroDaaS as an efficient, flexible and scalable UAV programming framework for autonomous aerial analytics.
\end{abstract}

\section{Introduction}
\label{sec:intro}

\paragraph*{Opportunities} 
Unmanned Aerial Vehicles (UAVs), commonly referred to as drones, equipped with onboard computing and communication capabilities are increasingly deployed as on-demand aerial services for applications such as disaster response~\cite{liu2023dome,10556878,KUMAR2024108977}
, precision agriculture~\cite{QU2024108543,betti2024drone}
, urban traffic management~\cite{10556942,BARMPOUNAKIS202050}, and accessibility~\cite{raj2023ocularone,avila2017dronenavigator}. These applications rely on deep learning-driven analytics, such as object detection and anomaly identification, using real-time camera feeds and sensor data for autonomous decision-making. 

\textit{Drones-as-a-Service (DaaS)} extends this paradigm by enabling UAVs to be accessed as scalable, autonomous services to support diverse domains. DaaS has been proposed for logistics, where drones optimize package delivery while reducing environmental impact~\cite{shahzaad2019composing}, and in healthcare, for timely transport of medical supply to disaster-affected areas~\cite{alkouz2020swarm,shahzaad2020game}. Advances in edge computing and networked architectures further enhance DaaS by enabling distributed processing and multi-drone coordination~\cite{chu2021holistic,li2024uav}. However, while these explore custom application design and runtime optimizations, there is an absence of a unified programming framework and runtime to intuitively compose and deploy such DaaS applications for heterogeneous drone platforms. 

\paragraph*{Challenges}
As drones gain popularity, startups, research groups, and enterprises are actively building drone-powered services across various domains. However, just as the true potential of the web was realized through the inception of service-oriented architectures (e.g., HTTP, REST, XML/JSON)~\cite{pautasso2008restful}, a similar transformation is required to rapidly design sophisticated drone-based applications in a vendor-agnostic, platform-neutral manner to accomplish Drones-as-a-Service (DaaS). Companies like DJI, Parrot, and Skydio offer diverse drone platforms, requiring \textit{proprietary} Software Development Kits (SDKs) such as DJI Mobile SDK and Parrot Ground SDK~\cite{dji_mobile_sdk} to program them. These offer APIs for \textit{low-level control} over drone navigation and on-board sensors.
Custom drone applications developed using these SDKs are limited to that vendor, and often even a specific drone model as the SDK capabilities differ. This prevents their portability to other drones and substantially increases developer effort. 

Further, these SDKs \textit{lack high-level primitives} for task-driven application composition. Practical DaaS applications require real-time \textit{analytics} for autonomous navigation and intelligence. Given the pervasiveness of Deep Neural Network (DNN) models over vision and sensor data, these analytics also require the efficient use of \textit{edge accelerators}, such as Nvidia Jetson, present onboard or co-located with the drone for rapid inferencing~\cite{uav_edgeinf}. These need to be complemented by abundant cloud resources and Inferencing-as-a-Service, intermittently accessible through cellular networks~\cite{raj2024adaptiveheuristicsschedulingdnn,he2021confect}. Also, the DaaS provider will require \textit{multi-tenancy}, using the same drone platform to host multiple applications that run concurrently and need to be sandboxed from each other~\cite{androne}.

Lastly, given the cyber-physical and safety-critical nature of drone applications, they must be tested in simulation environments like \textit{Gazebo} before their field deployment~\cite{1389727}. But the lack of interoperable and portable drone APIs creates high friction in development and testing -- much like the pre-iOS and pre-Android era, when mobile app and service development was fragmented and restricted to device manufacturers.

\paragraph*{Motivation} We identify two representative drone service paradigms that leverage a \textit{sensing-analytics-navigation} loop and pose key requirements for a DaaS programming framework:
(1) \textit{Waypoint-Driven Services (WDS)}: Drones follow a predefined trajectory based on waypoints, executing sensing or analytics tasks at specific locations or throughout the flight. E.g., UAVs can capture traffic images along a city road network to assess congestion and optimize traffic signaling~\cite{khan2017uav,khochareTON}. 
(2) \textit{Analytics-Driven Services (ADS)}: Drones autonomously adjust their trajectory based on real-time analytics. For instance, UAV-based assistive services can track a visually impaired individual by detecting a hazard vest via a DNN, issuing navigation commands, and providing obstacle alerts~\cite{raj2023ocularone}.
These emerging service models highlight the need for a service-oriented framework that offers intuitive interfaces 
to seamlessly compose applications based on these patterns. 

\paragraph*{Gaps}
The proprietary SDKs offered by commercial drone manufacturers are incompatible with others. While they offer niche consumer-facing features such as $360^{\circ}$ rotation to capture selfies, there are limited higher-order primitive to ease programming~\cite{dji_mobile_sdk}.
The open-source drone ecosystem has autopilot frameworks like PX4~\cite{px4_web_ref} and ArduPilot~\cite{ardupilot_web_ref} with extensible interfaces for UAV navigation. These are adopted by companies such as HolyBro and Intel Aero, as well as hobbyists for custom drone design. However, these frameworks operate at a low level of abstraction, offering SDKs limited to controller-level commands. They are also incompatible with proprietary drone platforms and lack high-level service primitives for analytics and edge-cloud operations.

Research efforts like AeroStack2\cite{fernandez2023aerostack2} and UAL\cite{real2020unmanned} provide hardware-layer abstractions for interoperability across drone platforms. While we leverage AeroStack2 as an enabling layer, these act more like ``device drivers'' and lack high-level service primitives for analytics and orchestration. Other works focus on Domain-Specific Languages (DSLs) for specific tasks like video sensing\cite{10.1145/3486607.3486750}, mission optimization\cite{10.1145/3386901.3388912}, or swarm coordination~\cite{10.1109/IROS.2016.7759558}, but they do not generalize to diverse drone applications. Our prior poster paper~\cite{10898177} demonstrates integration of DNN analytics but does not exploit the edge-cloud continuum or complex compositions.

\paragraph*{Contributions}
In this paper, we propose \textbf{AeroDaaS}, an interoperable DaaS framework that enables seamless composition of analytics-driven drone services. AeroDaaS provides a unified set of interfaces for sensing, analytics and navigation across drone platforms, abstracting the complexities of drone programming. By exposing these capabilities as modular, reusable APIs and data models, AeroDaaS lowers the barrier for designing and adoption of DaaS by drone fleet operators. 

The key contributions of this paper are:
\begin{enumerate}[leftmargin=*]
\item We discuss the \textit{key requirements} for AeroDaaS, motivated by a set of use-cases spanning diverse domains (\S~\ref{sec:use-cases}). 
\item We introduce the \textit{AeroDaaS architecture} (\S \ref{sec:arch}) and define the \textit{DaaS Programming Framework} (\S \ref{sec:prog-model}) that integrates sensing, navigation and analytics across edge-cloud continuum to concisely compose DaaS applications. 
\item We describe the \textit{runtime implementation} of AeroDaaS that supports multiple drone platforms, including the Gazebo simulator, and offers hooks for schedulers to execute analytics on edge accelerators and public clouds (\S \ref{sec:runtime}).
\item We evaluate AeroDaaS for two category of applications: VIP Tracking with situation awareness in the real-world using physical drones and Smart crop monitoring using a simulation environment to demonstrate the flexibility of our model (\S~\ref{sec:evals}). We also validate several edge--cloud schedulers that use Nvidia Jetson edge accelerators and AWS cloud for analytics.
\end{enumerate}
Besides these, we offer background (\S \ref{sec:bg}), review related work (\S \ref{sec:related}), and offer our conclusions (\S \ref{sec:conclusions}).

\section{Background}
\label{sec:bg}
This section presents key concepts and open-source drone toolchains used by \textit{AeroDaaS}, highlighting their interactions.

\subsection{Components in a Drone Ecosystem}
\paragraph*{Flight Controller}
This is the core hardware managing flight operations, processing sensor data, controlling motors, and handling payloads. Optimized for real-time, low-latency tasks, it ensures stability while delegating higher-level functions to an onboard computer. In AeroDaaS, AeroStack2 serves as the hardware abstraction layer, interfacing with Pixhawk~\cite{5980229} for flight control and sensor data.

\paragraph*{Companion Mission Computer} 
Advancements in System-on-Chip (SoC) and GPU-accelerated edge platforms like NVIDIA Jetson enable companion computers to efficiently handle computationally intensive tasks, such as DNN inferencing, offloaded from the flight controller. They run standard Linux OS, and may be on-board or co-located with the drone's base station via wireless links. AeroDaaS abstracts the mission computer to include edge and/or cloud resources and enables high-level analytics, navigation and orchestration, complementing the flight controller.

\paragraph*{AutoPilot} The autopilot software interfaces with the flight controller to execute high-level commands like waypoint navigation, flight mode selection, and autonomous decision-making. Open-source platforms like PX4~\cite{px4_web_ref} and ArduPilot~\cite{ardupilot_web_ref} enable precise flight control and seamless integration with onboard systems. AeroDaaS abstracts and extends these to integrate analytics-driven autonomous navigation.

\paragraph*{Offboard APIs} Offboard APIs provide programmatic control over drones, supporting high-level languages like C++ and Python for autonomous operations and cloud integration. They enable features for navigation and sensor data access. In AeroDaaS, these APIs connect drones to external compute resources and sensors, facilitating seamless analytics and decision-making.

\subsection{Robot Operating System (ROS2)}
Robot Operating System 2 (ROS 2) is an open-source framework that facilitates the development of modular and scalable robotic systems, with tools for perception, navigation and control \cite{doi:10.1126/scirobotics.abm6074}. It follows a publish-subscribe architecture using nodes, topics and standard message formats for seamless data exchange. AeroDaaS leverages ROS 2's extensive ecosystem, enabling interoperability with diverse applications. Additionally, ROS 2 supports integration with simulators like Gazebo~\cite{1389727} using standard plugins, allowing for efficient testing and validation in simulated environments.

\subsection{Aerostack2}
Aerostack2 (AS2)~\cite{fernandez2023aerostack2} is an open-source framework built on ROS 2, designed to simplify the development of aerial robotics systems. It allows developers to define mission plans by specifying UAV tasks and supports heterogeneous drones, including beginner-friendly models like DJI Ryze Tello, open-source controllers like PX4, and proprietary DJI drones via SDKs. AS2 ensures interoperability through a common API and a plugin-based design, allowing integration with new hardware by implementing backend support for different platforms. Given its comprehensive ROS2 functionalities, we leverage Aerostack2 as our hardware abstraction layer, enabling flexible and scalable UAV deployments.

\section{Motivation and Requirements for DaaS}
\label{sec:use-cases}

Next, we describe representative and diverse analytics-driven drone applications from real-world domains, which we use to synthesize the requirements for a service-oriented and cross-platform DaaS programming framework.

\subsection{VIP Navigation and Situation Awareness}
Drones can serve as an assistive technology for differently-abled individuals, offering real-time situational awareness and navigation support. Ocularone~\cite{raj2023ocularone} provides a drone-based service where a buddy drone autonomously follows a Visually Impaired Person (VIP) in urban spaces. The drone detects the VIP wearing a hazard vest using a Yolov8-based \textit{Hazard Vest (HV) detection model} running on live video feeds. These detections are used by a tracking algorithm to generate drone navigation signals for safe and consistent following of the VIP. Additionally, \textit{Body Pose Estimation}, powered by a ResNet18-based DNN and an SVM classifier, enhances situational awareness by detecting critical events like falls. \textit{Distance estimation} and \textit{obstacle detection analytics} further enable real-time alerts and autonomous navigation.

\subsection{Smart Crop Monitoring} 
Drones play a crucial role in agricultural technology by enabling efficient crop monitoring across vast rural farmlands, where manual surveys are time-intensive~\cite{betti2024drone}. We outline two types of survey-based drone services: coarse and fine grained monitoring. In coarse grained surveys, drones follow predefined GPS waypoints from a higher altitude, capturing aerial footage to identify farm activities, such as detecting tractors, machinery or anomalies in field coverage. In contrast, fine-grained surveys focus on detailed imagery of crops, leaves, fruits, and soil to detect pests, assess plant health and determine irrigation or pest-control needs. Advanced analytics, in real-time or offline, process these images for insights. Analysis of soil moisture data collected from ground sensors by the drones further enhances precision agriculture.

\subsection{Wildfire Management}
During active wildfires, drones provide critical real-time data on fire intensity, direction and spread, assisting emergency teams in containment planning~\cite{liu2023dome}. Drones survey fire-prone regions using a fire detection model based on YOLOv11, trained on fire datasets such as FlameVision~\cite{ibn_jafar_2023_flamevision}, to identify fire in video footage and map active fire zones. These enable continuous fire tracking and adaptive trajectory adjustments to avoid hazardous areas. Post-wildfire, drones facilitate damage assessment, aiding government authorities in evaluating the impact and planning recovery efforts.

\subsection{Requirements}\label{sec:requirements}
We highlight key requirements based on these and other analytics-driven aerial applications. 

\subsubsection{Composable Interfaces for Common Application Logic}
A standardized set of high-level programming interfaces is essential for addressing common drone application needs. While existing solutions provide limited support based on specific drone hardware, they lack cross-platform compatibility. E.g., DJI SDKs offer mission-specific services such as \textit{Revolve}, \textit{Follow} and waypoint-based navigation, but these do not extend to other drones. We require a unified framework with APIs for \textit{sensing, navigation and analytics}, enabling seamless integration across diverse drone hardware. These allow developers to build and extend drone application services while incorporating automation, real-time decision-making and platform interoperability.

\subsubsection{Hardware-Agnostic Implementation}
A unified software stack is essential for seamless drone application development across diverse hardware platforms. To ensure interoperability, a service-oriented architecture should integrate \textit{sensing, navigation, and analytics} as core services, enabling real-time data collection and dynamic control. Additionally, robust development demands simulation environments like Gazebo~\cite{1389727}, allowing for thorough testing and validation in both virtual and real-world scenarios without modifying application logic.

\subsubsection{Analytics Services across Edge-Cloud Continuum}
Services for executing analytics and making autonomous decisions are essential for drone applications. Given the computational demands of DNN-based analytics -- often operating on high-bandwidth sensor data and requiring concurrent execution -- resource constraints on the mission computers (edge devices) can limit performance. Applications must be able to define priorities and performance guarantees for their analytics, ensuring critical tasks meet latency and throughput needs. This requires the ability to seamlessly integrate intelligent schedulers of analytics across inferencing services operating on the \textit{edge-cloud service continuum}, to ensure efficient workload execution to meet QoS and QoE needs~\cite{chen2019uav,raj2024adaptiveheuristicsschedulingdnn}. 

\subsubsection{Service Deployment and Execution Isolation}
Drone applications often require multiple external libraries and DNN models, which may have conflicting dependencies. To ensure seamless deployment across diverse compute environments while maintaining system stability, we require a containerized service model, such as AnDrone~\cite{van2019androne} operates virtual drones in isolated cloud environments. Encapsulating applications and their inferencing services as containers allows on-the-fly installation, prevents dependency conflicts and ensures execution isolation, enhancing security and resource efficiency~\cite{10487048}. Lastly, the proposed programming framework should be versatile enough to implement a broad range of application and service scenarios. While the current work focuses on single-drone deployments, in future, this should consider concurrent applications and multi-drone fleet integration.

\section{AeroDaaS Architecture and Programming Framework}\label{sec:prog-model}
This section outlines the high level design and core programming APIs and data models provided by AeroDaaS, enabling developers to intuitively compose DaaS applications. 

\subsection{AeroDaaS Architecture}\label{sec:arch}
\begin{figure}[t]
\vspace{-0.1in}
\centering
\includegraphics[width=0.8\columnwidth]{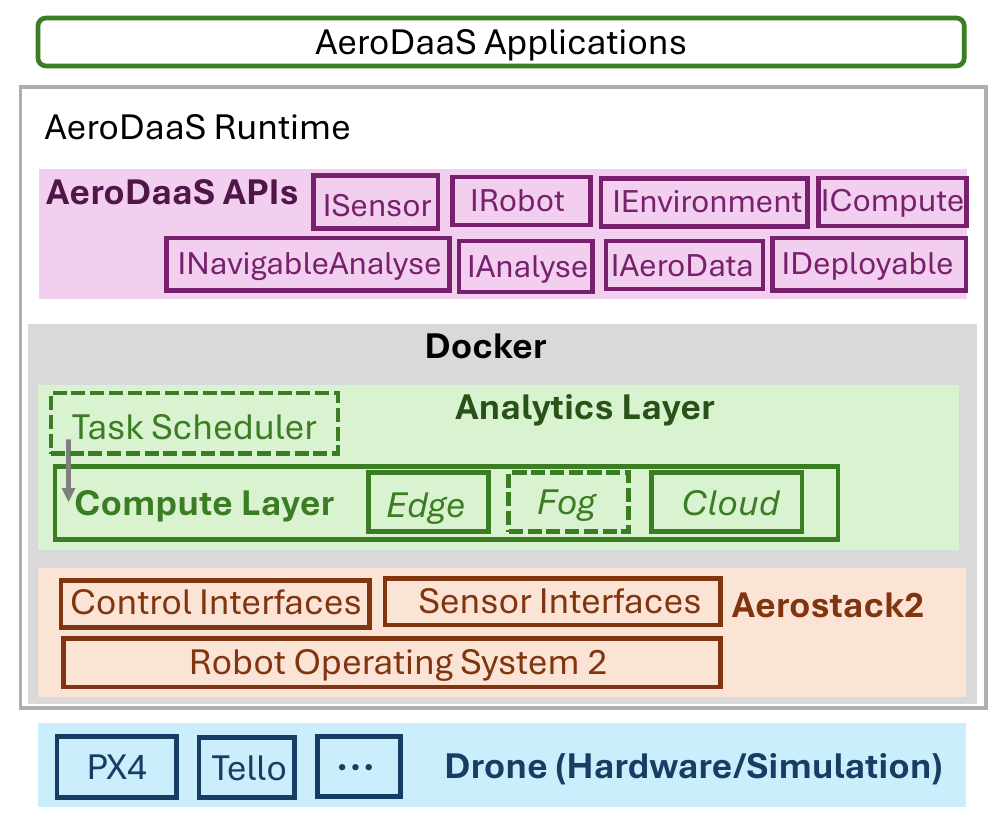}
\caption{AeroDaaS Architecture and Services} 
\vspace{-0.15in}
\label{fig:architecture}
\end{figure}

Fig.~\ref{fig:architecture} shows the AeroDaaS architecture. DaaS applications are built using AeroDaaS's API, which abstracts low-level drone navigation, controls, sensing and analytics into intuitive, composable Python-based interfaces. Internally, some of these APIs extend Aerostack2’s capabilities by integrating with onboard sensors and actuators across both real-world and simulated environments.
A key novelty in AeroDaaS is its analytics service layer, which enables DNN-based models to run across edge, fog and cloud resources for a hybrid execution model. These analytics services can be intelligently used by plugging in task schedulers to dynamically distribute workloads to balance performance and cost.
To ensure modularity and isolation, AeroDaaS employs containerized service execution using Docker. Each service, such as sensor access, analytics and compute management, runs in an independent container, allowing seamless deployment and encapsulation between layers. For instance, the \textit{ISensor} Service interfaces with Aerostack2’s sensor layer, while the \textit{IDeployable} Service orchestrates workloads across edge and cloud compute resources within the Compute Layer.

\begin{figure}
\vspace{-0.1in}
  \centering
  \includegraphics[width=1\columnwidth]{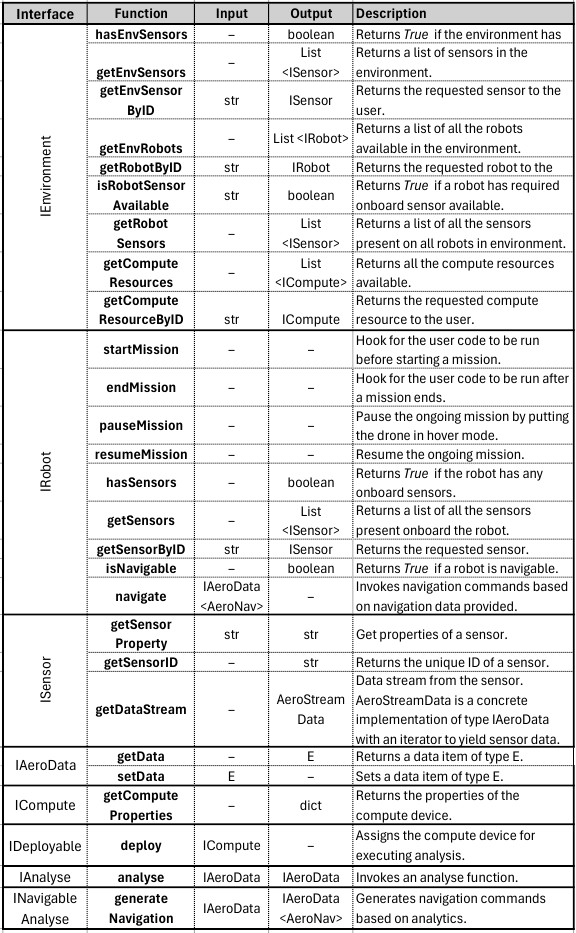}
  \caption{APIs provided by AeroDaaS}
\end{figure}

\subsection{Environment Management Interface (\textit{IEnvironment})}
The \textit{IEnvironment} interface provides services for accessing environmental resources, including sensors, robots (drones) and computing devices. The details about the drones, sensors and edge/cloud computing devices in the \textit{IEnvironment} is bootstrapped from a configuration file or, later, can be auto-discovered dynamically. Developers can query the availability of sensors using \textit{hasEnvSensors()}, retrieve all sensors via \textit{getEnvSensors()}, or fetch a specific sensor using \textit{getEnvSensorByID(sID)}. These may be sensors in the neighborhood of the drone or on-board the drone. Similarly, \textit{getEnvRobots()} and \textit{getRobotByID(rID)} allow users to list and retrieve robots (drones) present in the environment. Sensors that are present within the drone can be accessed using similar APIs, \textit{isRobotSensorAvailable(sID)} and \textit{getRobotSensors()}. 

Additionally, compute resources that are accessible within the environment can be discovered through \textit{getComputeResources()}, and specific ones can be fetched using \textit{getComputeResourceByID(cID)}. These again may be edge devices on-board the drone, nearby edge accelerators or remote cloud resources. These services ensure that developers can dynamically interact with environmental elements, facilitating adaptive and responsive drone applications. While we implement some of these in our current work, building AeroDaaS to tap into a drone service provider's resources by using their URL to initialize the \textit{IEnvironment} with their resources instead of local resources, is one of our future works.

\subsection{Mission Control Interface (\textit{IRobot})}
The \textit{IRobot} interface provides mission execution and control for managing drone operations. The \textit{startMission()} and \textit{endMission()} interfaces serve as hooks for user-defined logic before and after mission execution, respectively. Missions can be temporarily halted using \textit{pauseMission()}, allowing the drone to enter a hover/standby state, and later resumed via \textit{resumeMission()}. Navigation-related services include \textit{isNavigable()}, which determines whether a robot supports navigation, and \textit{navigate(IAeroData<AeroNavigation>)} to execute navigation commands based on provided parameters. In future, this can even be extended to ground robots alongside UAVs. These services enable a hardware agnostic control and execution of drone missions.

\subsection{Sensor Data Access Services (ISensor)}
The \textit{ISensor} interface facilitates sensor data acquisition and management, enabling applications to interact with onboard and environmental sensors. The \textit{getSensorProperty(prop)} method allows querying specific sensor attributes, while \textit{getSensorID()} returns the unique identifier of a sensor. Real-time data streaming is supported through \textit{getDataStream()}, which provides continuous data from a sensor in the form of \textit{AeroStreamData}. This is further divided into \textit{IPushSensor} and \textit{IPullSensor} based on whether the user wishes to subscribe to events through a callback (push), or poll for new events periodically.

\subsection{Data Management Interface (IAeroData)}
The \textit{IAeroData} interface defines standard interfaces for managing and manipulating data items within AeroDaaS. It introduces a generic data type, \textit{AeroData<E>}, which encapsulates a single data item of type \textit{E}. This enables seamless handling of diverse data types such as images, sensor readings and analysis outputs, by setting the relevant type for \textit{E}. The \textit{getData()} service retrieves the encapsulated data, while \textit{setData(E data)} allows modifying its value. These ensure that AeroDaaS applications have standard mechanisms to access data and compose dataflow applications, with type specializations supporting both structured, unstructured and custom data formats. \textit{AeroStreamData} is another type of \textit{IAeroData} where a continuous stream of data items are made available to the applications, and is used by \textit{ISensor}.
The \textit{AeroListData} extension offers a list of items (rather than a stream), and is used with the \textit{AeroNavigation} data type to specify navigation waypoints or GPS coordinates for the drones.

\subsection{Compute Resource Management Interface (ICompute \& IDeployable)}
The \textit{ICompute} interface provides an interface to manage compute resources across edge and cloud, and initialize/access analytics on them. Developers can query the compute device's capabilities using \textit{getComputeProperties()}, which returns properties of the resource, such as processing power, accelerators, available memory, etc. Further, analytics services can be deployed on the \textit{ICompute} using \textit{IDeployable}, which can encapsulate an edge resource, a cloud resource, or a scheduler that encapsulates edge and cloud resources and load-balances across them. The framework can be easily extended to support different type of compute resources and schedulers, as we demonstrate.

\subsection{Analytics and Analytics-based Navigation (IAnalyse \& INavigableAnalyse)}
The \textit{IAnalyse} interface enable analytics-driven decision-making for drone applications. It allows invoking an analytics function \textit{analyse(IAeroData)} on a data specified by the application to support AI-driven insights for mission planning and execution, e.g., performing object detection and returning bounding boxes over a video stream input. 
Further, \textit{INavigableAnalyse} is a specialization that returns an analytics output of type, \textit{AeroNavigation} when invoking \textit{generateNavigation(AeroData)}. So passing an sensor event stream, e.g., current location, can have the analytics return \textit{AeroData<AeroNavigation>} that has a series of navigation instructions that can be used to control the trajectory drone. These services allow AeroDaaS applications to integrate DNN models and advanced computational techniques for real-time analysis, enhancing autonomy and operational efficiency in drone missions.
Users can also implement custom analytics. E.g., we have \textit{MonitoringAnalytics} that implements \textit{IAnalyse} and can monitor and visualize battery and odometry sensors.

\begin{figure*}[t!]
\vspace{-0.1in}
  \centering
  \includegraphics[width=1\textwidth]{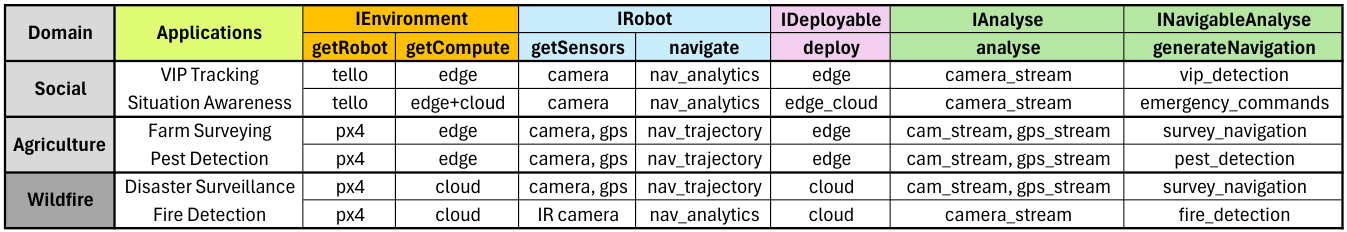}
    \vspace{-0.25in}
  \caption{Diverse applications composed using various interfaces offered by AeroDaaS  }
  \label{fig:applications-primitives-table}  
    \vspace{-0.15in}
\end{figure*}

\section{Developing Applications using AeroDaaS}\label{sec:application-psuedocodes}
In this section, we use the AeroDaaS APIs to to compose the DaaS applications from our motivating use-cases. We provide representative code snippets that use the programming and data interfaces to compose application services as dataflows. Fig.~\ref{fig:applications-primitives-table} illustrates the coverage of our APIs by the applications. The environment initialization bootstraps the resources available to the developer. For simplicity, we assume obstacle avoidance is enabled natively by default for the drones.

\subsection{VIP Navigation and Situation Awareness}\label{sec:application-psuedocodes:vip}
Here, we showcase an analytics-driven application where the drone's camera stream is passed to a Hazard Vest detection analytics, whose bounding box output is used to generate navigation commands by using a PID controller to help track the VIP. The same camera stream is also passed to a Body Pose analytics to detect specific gestures by the VIP, that can then be used to, e.g., have the drone land if the VIP raises their hand, or detect a fall situation.

\begin{lstlisting}[language=python]
env: IEnvironment = Environment("config.json")
# Get Edge only compute and Edge+Cloud scheduler
edge_compute: ICompute = env
    .getComputeResourceById("edge")
edge_cloud_compute: ICompute = env
    .getComputeResourceById("edge_cloud")
# Initialize tello drone and start mission
drone: IRobot = env.getRobotById("tello")
drone.start_mission()
# Get the camera data stream
camera: IPushSensor[Image] = drone
    .getSensorById("camera")
camera_data_stream: AeroStreamData[Image] = 
    camera.getDataStream()
# Detect Visually Impaired Person based on Ocularone 
# analytics over images, deployed to edge compute
vip_detection_analytics: IAnalyse = VipDetectionAnalytics()
vip_detection_analytics.deploy(edge_compute)
vip_detection_data_stream: AeroStreamData[Bbox] = 
    vip_detection_analytics.analyse(camera_data_stream)
# Pass the detected object's bounding box data 
# to the navigation path generation analytics
follow_navigation_analytics: INavigableAnalyse = FollowObjectNavigation()
follow_navigation_analytics.deploy(edge_compute)
navigation_data: AeroStreamData[AeroNavigation] = follow_navigation_analytics
    .generate_navigation(vip_detection_data_stream)
# Emergency commands analytics deployed to a
# cloud + edge compute resource
emergency_commands_analytics: INavigableAnalyse = BodyPoseAnalytics()
emergency_commands_analytics.deploy(edge_cloud_compute)
# Subscribes to the camera stream to generate 
# emergency commands if required
emergency_commands_data: AeroStreamData[AeroNavigation] = 
    emergency_commands_analytics.generate_navigation(camera_data_stream)
# Combines the navigation commands from both the
# analytics 
drone.navigate([navigation_data, emergency_commands_data])
drone.endMission()
\end{lstlisting}

\subsection{Smart Crop Monitoring} \label{sec:application-psuedocodes:crop}
We describe a waypoint-driven application where the length and width of the required survey area is provided by the user, and commands are generated to cover that. The camera feed is shared with another downstream application, such as saving it to a file or publishing to a topic. 

\begin{lstlisting}[language=python]
# Using a drone service provider (Future Work) 
# besides local edge resources 
env: IEnvironment = Environment("AeroDaaS://droneserviceprovider.com/foo/bar/config.json")
edge_compute: ICompute = env
    .getComputeResourceById("edge")
# Select a PX4 flight controller-based drone and start
drone: IRobot = env.getRobotById("px4")
drone.start_mission()
# Get the GPS and Camera data streams
gps: IPushSensor[LatLong] = drone
    .getSensorById("gps")
gps_data_stream: AeroStreamData[LatLong] = 
    gps.getDataStream()
camera: IPushSensor[Image] = drone
    .getSensorById("camera")
camera_data_stream: AeroStreamData[Image] = 
    camera.getDataStream()
# Write gps sensor data to a file
save_gps_analytics: IAnalyse = 
    SaveDataAnalytics("/path/to/directory")
save_gps_analytics.deploy(edge_compute)
save_gps_analytics.analyse(gps_data_stream)
# Write camera sensor data to a file
save_camera_analytics: IAnalyse = 
    SaveDataAnalytics("/path/to/directory")
save_camera_analytics.deploy(edge_compute)
save_camera_analytics.analyse(camera_data_stream)
# Survey a 30m x 60m area from a height of 10m
nav_trajectory_data: AeroListData[AeroNavigation] = NavigationPathGenerator
    .get_rectangular_survey_path(
    length=30, width=60, height=10)
# Log waypoints given to the drone and actual 
# trajectory that is followed. MonitoringAnalytics
# automatically creates files for the metrics
odometry: IPushSensor = drone.getSensorById("odom")
odometry_data_stream: AeroStreamData[Odom] = odometry.getDataStream()
odometry_analytics: IAnalyse = MonitoringAnalytics(["battery", "trajectory"])
odometry_analytics.deploy(edge_compute)
_: AeroStreamData[Odom] = odometry_analytics.analyse(odometry_data_stream)
# Use the navigation commands to control the drone
drone.navigate(nav_trajectory_data)
# Complete waypoint-driven application
drone.end_mission()
\end{lstlisting}

\subsection{Wildfire Management} This is an analytics driven application where drone's trajectory is revisited based on the vision analytics running onboard the drone. Here, we highlight the unique features of AeroDaaS used and omit common code.

\begin{lstlisting}[language=python]
...
# Using the AWS cloud compute 
cloud_compute: ICompute = env
    .getComputeResourceById("cloud")
# Using px4 drone
drone: IRobot = env.getRobotById("px4")
# Deploy Fire detection on AWS, consume camera feed
infrared_camera: IPushSensor[Image] = ...
camera_data_stream: AeroStreamData[Image] = ...
fire_detect_analytics: IAnalyse = FireDetectionAnalytics()
fire_detect_analytics.deploy(cloud_compute)
fire_data_stream: AeroStreamData[Bbox] = 
  fire_detect_analytics.analyse(camera_data_stream)
# Alerts for detected fire
fire_alert_analytics: IAnalyse = 
    FireAlertAnalytics()
fire_alert_analytics.deploy(cloud_compute)
_ = fire_alert_analytics.analyse(fire_data_stream)
# Take evasive actions or raise alerts...
...
\end{lstlisting}

\section{AeroDaaS Runtime Implementation}
\label{sec:runtime}
\begin{figure}[t]
\vspace{-0.1in}
    \centering
    \includegraphics[width=0.8\columnwidth]{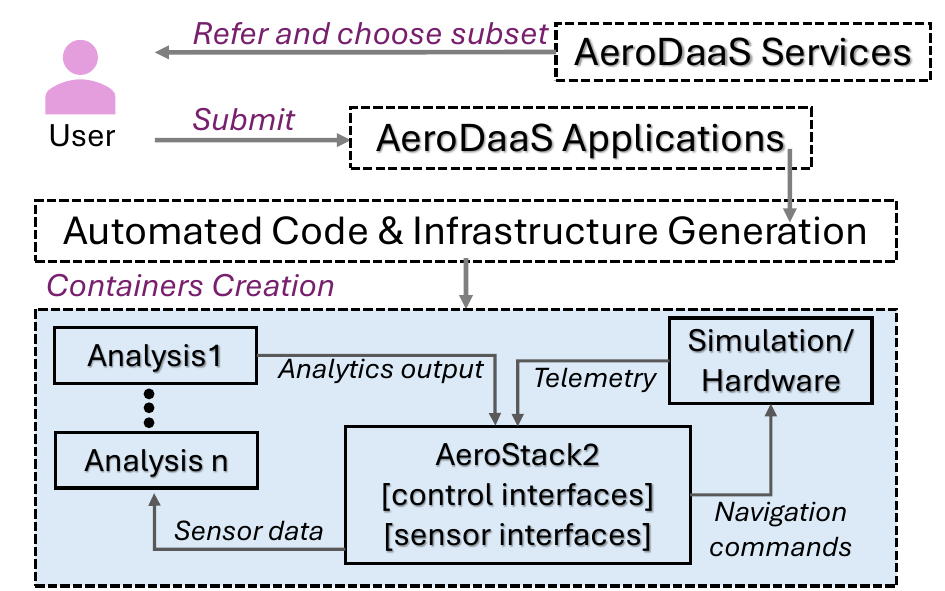}
    \caption{
    AeroDaaS Runtime Implementation}
    \label{fig:runtime-implementation}
    \vspace{-0.15in}
\end{figure}

Fig.~\ref{fig:runtime-implementation} illustrates the key components of the AeroDaaS runtime implementation. The user interacts with the AeroDaaS APIs, selects a subset of available functions to define their application, and submits their application through a user-facing file. This serves as the entry point for the automated code and infrastructure generation process. This is responsible for generating a Python-based ROS2 node for the main application that interacts with sensors and the drone hardware through Aerostack2 platform. Along with that, a set of bash files needed by AeroStack2 are generated to run the Docker containers for the ML models and the root Docker container to interact with them. Once the necessary files are generated, multiple containers are launched at runtime to execute the applications. The following subsections details each step.

\subsection{Automated Code and Infrastructure Generation}
We implement our runtime in \textit{Python3} and provide equivalent Python APIs for the AeroDaaS services. Users develop applications as Python scripts, which execute within an isolated container. This container is set up using a pre-configured Docker image that includes (a) the AeroStack2 library, (b) backend support for the target drone hardware, and (c) the user’s transpiled application code embedded in a ROS2 node with sensor subscriptions and analytics interactions. The Docker image comes with all necessary dependencies pre-installed. For code generation, we use Jinja2~\cite{jinja2}, a fast and extensible templating engine that supports Python-like syntax with special placeholders.

AeroDaaS relies on Aerostack2 plugins as the hardware abstraction layer, requiring them to be active before any interface is used. When a user executes their Python application, it first invokes \textit{generate\_code}, the core function responsible for infrastructure setup. This function generates a Python-based ROS2 node file containing services for analytics, navigation, sensing, and compute, along with corresponding bash scripts to launch necessary services. The main file is dynamically populated based on user inputs. The bash scripts then initialize the required containers, including Aerostack2 services, analytics modules, and optional simulation environments if no physical drone is used. Finally, the generated ROS2 node executes, starting the mission.

\subsection{Containers Coordination using AeroDaaS Services}
AeroStack2 provides \textit{control interfaces} for drone navigation and \textit{sensor interfaces} for accessing onboard data, both exposed via ROS topics and messages. AeroDaaS ensures seamless inter-container communication by running all containers on the \textit{host} network, enabling data exchange over \textit{localhost}. The \textit{ISensor} service allows analysis containers to retrieve sensor data, while \textit{IAnalysis} facilitates the transfer of analytics outputs to the AeroStack2 container. When a simulation is enabled, \textit{INavigate} manages navigation commands for Gazebo, and \textit{IAeroData} ensures telemetry data is transmitted efficiently to AeroStack2.

\section{Evaluation}
\label{sec:evals}
We perform a detailed evaluation of AeroDaaS for two categories of the motivating applications that we have implemented using our programming framework and deployed within our runtime. These validate a real-world deployment for the \textit{VIP Tracking and Situation Awareness} application using the Tello Drone using hardware-in-the-loop (HITL), and a simulation-based evaluation of \textit{Farm Surveying and Pest Detection} using a X500 drone using software-in-the-loop (SITL). These demonstrate the flexibility of the framework to execute the application for different target platforms. We also evaluate the use of edge and cloud resources individually, and combined together using a sophisticated scheduler, for executing the analytics that drive the DaaS application.

\begin{figure}[!t]
  \centering
    \begin{minipage}[t!]{0.43\columnwidth}
    \centering
    \includegraphics[width=0.8\textwidth]{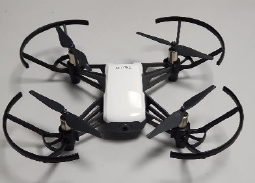}
   \label{fig:dji_tello_hardware}
   \\
   \includegraphics[width=0.8\textwidth]{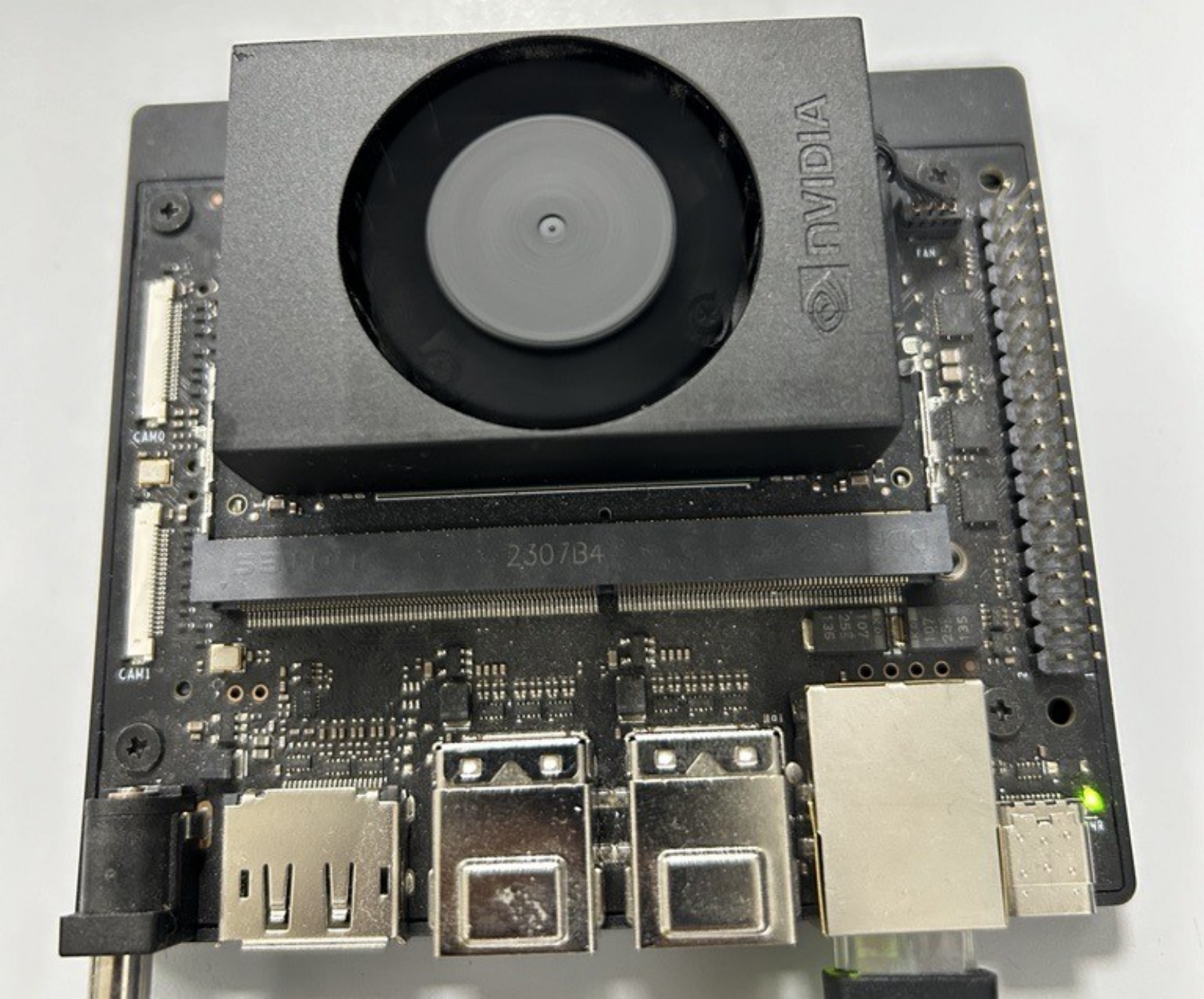}
    \caption{DJI Tello Quadcopter (top) and Nvidia Jetson Orin Nano (bottom)}
    \label{fig:orin_hardware}
\end{minipage}\quad
  \begin{minipage}[t!]{0.5\columnwidth}
  \centering
    \includegraphics[width=0.8\textwidth]{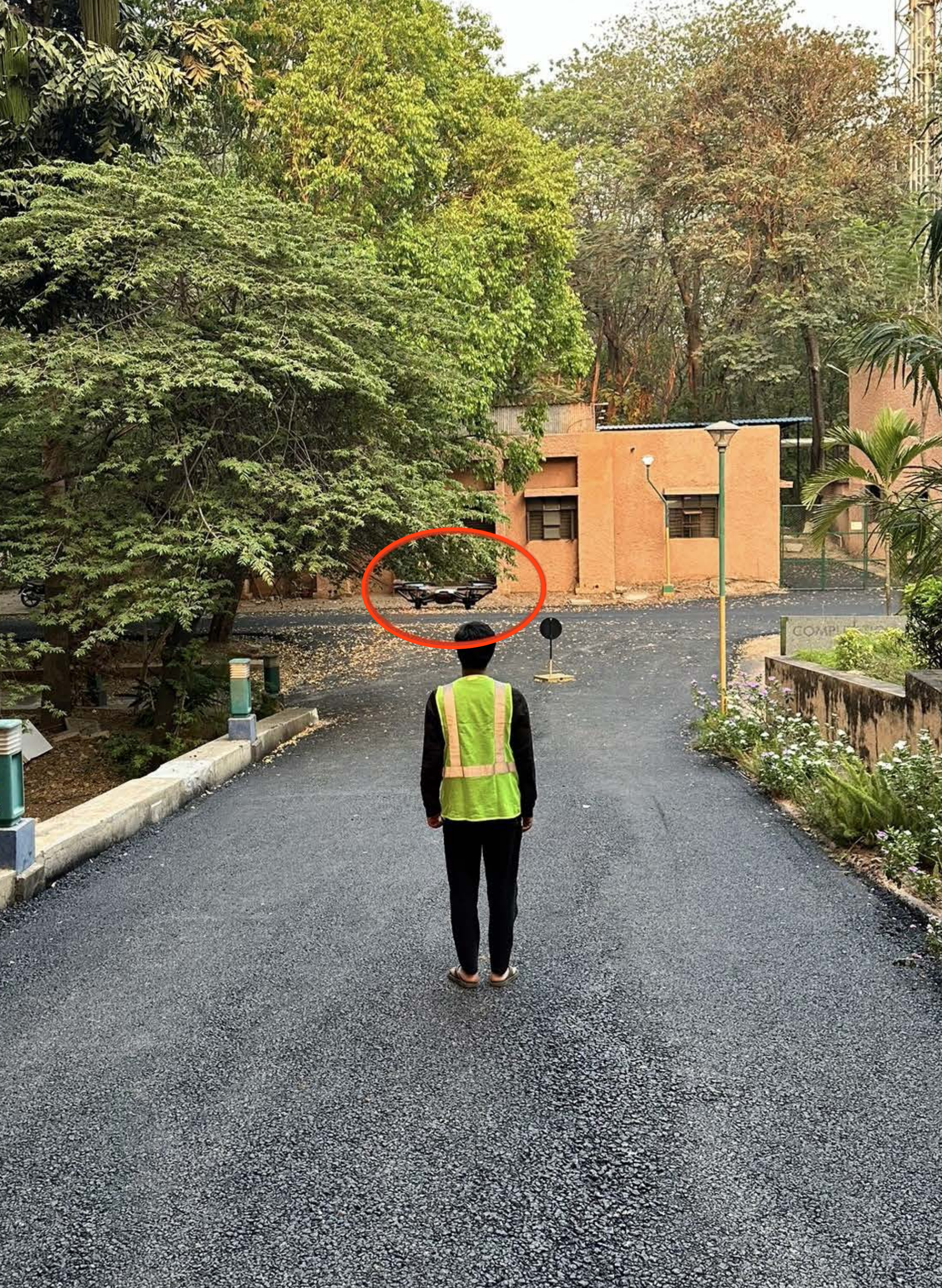}
    \caption{VIP tracking application on DJI Tello UAV (red circle)}
    \label{fig:hardware_tracking}
    \end{minipage}
\vspace{-0.15in}
\end{figure}

\subsection{Setup}
For the HITL experiment, we use a DJI Ryze Tello drone and an Nvidia Jetson Orin Nano edge accelerator (Fig.~\ref{fig:orin_hardware}). The Orin Nano has a six-core Arm Cortex-A78AE CPU, 1024 Ampere CUDA cores, and 8GB of RAM shared by CPU and GPU. It has a power range of $7$--$15$W, is powered by a portable power bank, has compact dimensions of $100 \times 79$mm, and connects to the drone through WiFi. The DJI Tello is equipped with an onboard camera capable of generating live video feeds at 30 FPS at $720p$ resolution. Fig.~\ref{fig:hardware_tracking} shows a third-person view of the experiment where Tello drone (red circle) is following a person wearing a hazard vest from a safe distance based on the application logic. The Tello is connected to the Orin Nano kept nearby using WiFi. 

For the \textit{VIP Tracking} application, we deploy a YOLOv8 (nano) DNN model trained to detect a fluorescent hazard vest, combined with a PID control loop analytics for the drone to following the VIP based on the bounding boxes (Sec.~\ref{sec:application-psuedocodes:vip}). This AeroDaaS application uses an analytics-based navigation. The \textit{Situation Awareness} application extends this with a body pose estimation DNN and a classifier to detect a \textit{fall} position. AeroStack2 and the required DNN models are containerized using Docker and run on the Orin Nano for edge-based experiments, while AWS Lambda is utilized for cloud-based experiments. We follow the cloud experiment setup outlined in~\cite{raj2024adaptiveheuristicsschedulingdnn}. 

For SITL simulations, we use the Gazebo simulator (GZ Garden) integrated with ROS2 (Humble) running inside a Docker container on a GPU workstation equipped with an AMD Ryzen 9 3900X 12-core CPU, 24GB RAM, and an Nvidia RTX 3090 GPU. The \textit{Farm Surveying and Pest Detection} application (Sec.~\ref{sec:application-psuedocodes:crop}) is evaluated using the \textit{baylands} virtual world in Gazebo, with an \textit{X500} drone featuring onboard GPS. The survey parameters define a rectangular area of $60m \times 30m$ at an altitude of $10$m from the starting position, with the drone flying at a speed of $1m/s$. This AeroDaaS application uses a waypoint based navigation.

\begin{figure}[!t]
    \centering
    \subfloat[DNN Models on Edge]{%
    \includegraphics[width=0.49\linewidth]{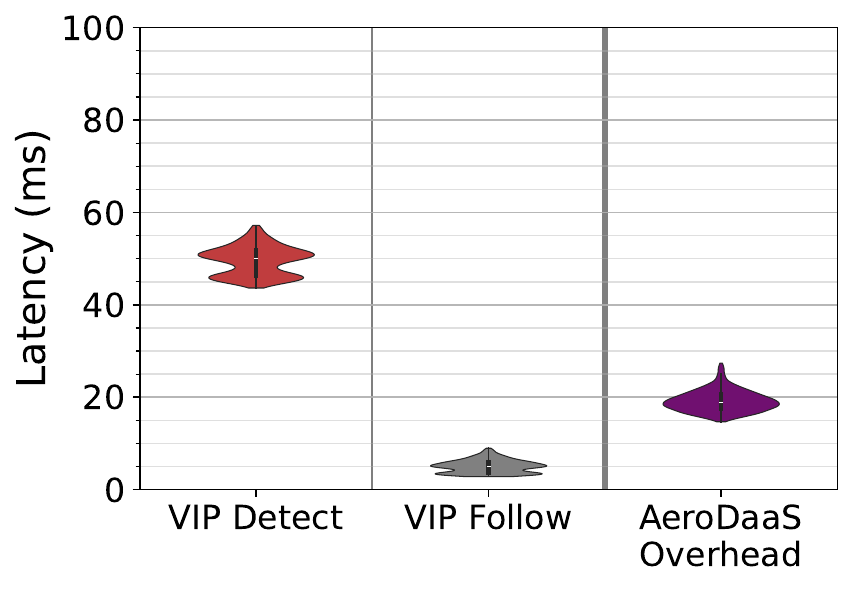}%
    \label{fig:edge-vip-tracking}%
    }~
    \subfloat[DNN Models on Cloud]{%
    \includegraphics[width=0.49\linewidth]{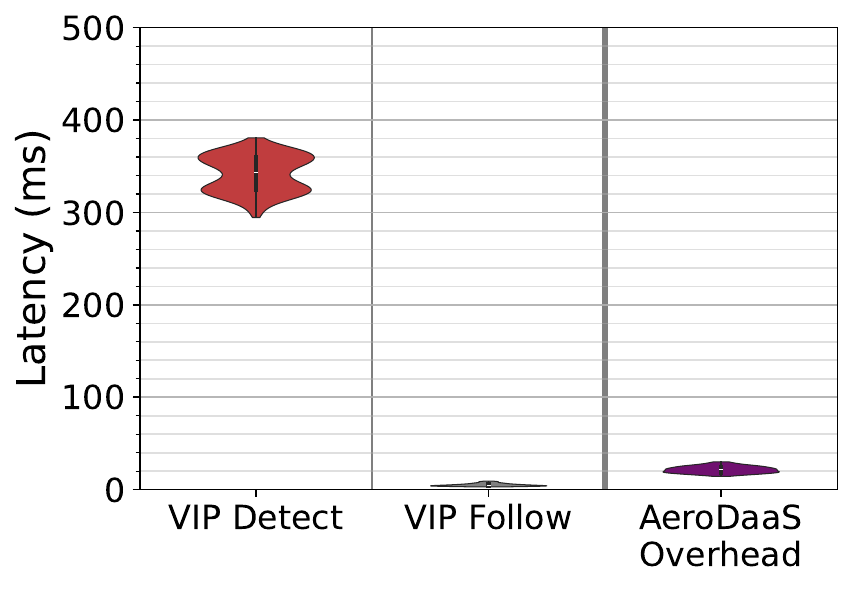}%
    \label{fig:cloud-vip-tracking}%
    }
\caption{End-to-end AeroDaaS and DNN inference latency for VIP Tracking application on hardware.}
\label{fig:vip-tracking}
\vspace{-0.15in}
\end{figure}

\begin{figure}
    \centering
    \includegraphics[width=0.7\linewidth]{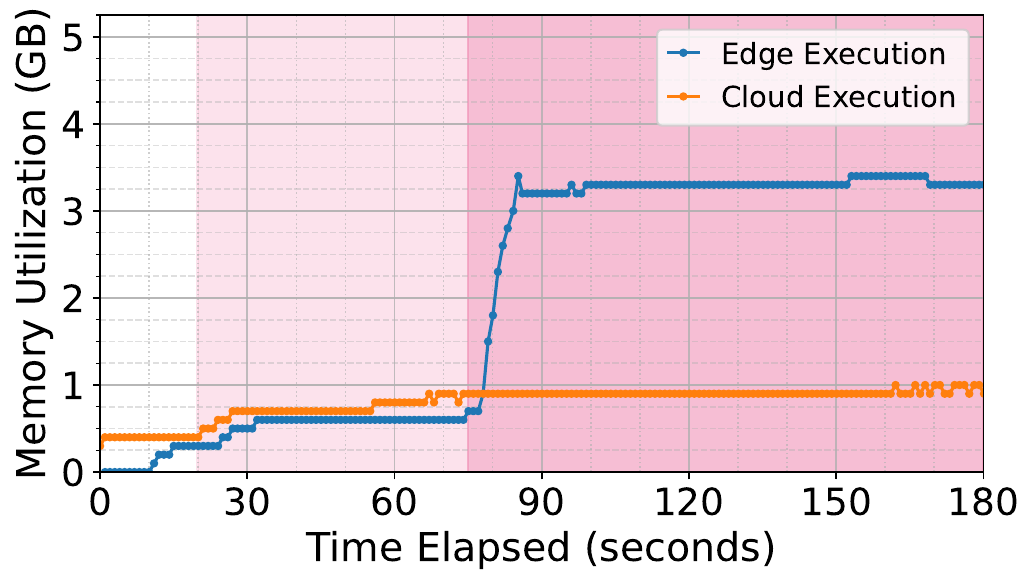}
    \vspace{-0.07in}
    \caption{AeroDaaS memory utilization on Jetson Orin Nano}
    \label{fig:system-memory}
    \vspace{-0.15in}
\end{figure}

\subsection{Results}
Next, we present key insights from our experiments that validate these two applications that are implemented and orchestrated using AeroDaaS, each running for over 120 seconds. 

\subsubsection{\hilite{The AeroDaaS runtime is lightweight and has a minimal overhead of $\approx20$ms and $\leq0.5$GB RAM usage on edge-accelerated hardware}}
In the hardware experiments, we analyze the overhead latency of AeroDaaS, as shown in Fig.\ref{fig:vip-tracking}. The end-to-end latency includes receiving image frames from the drone over WiFi, performing DNN inference on either the edge device (Fig.\ref{fig:edge-vip-tracking}) or the cloud (Fig.~\ref{fig:cloud-vip-tracking}), denoted as \textit{VIP Detect}, processing the results, generating navigation commands using a PID control loop (\textit{VIP Follow}), and transmitting these commands to the drone. The analysis runs on every alternate frame, i.e., at $15$ FPS. AeroDaaS overhead is measured as the end-to-end latency excluding \textit{VIP Detect} and \textit{VIP Follow}. We observe that DNN inference takes approximately $50$ ms on the edge and $325$ ms on the cloud, while \textit{VIP Follow}, which involves simple mathematical computations, executes within $10$ ms. Overall, AeroDaaS introduces an overhead of only $\approx20$ ms out of $80$ms and $355$ms end to end latency on the edge and cloud respectively, per frame on the Jetson Orin Nano, a modest trade-off for the flexibility it provides.

Furthermore, we analyze the memory utilization of AeroDaaS to ensure that its overhead remains minimal. As shown in the Fig.~\ref{fig:system-memory}, AeroDaaS starts up at $t=25$s, and the analytics services begin computation at $t=75$s, both of which are highlighted by the shaded regions. Since the models run on the edge device, we observe a noticeable increase in memory usage after reaching approximately $3.5$~GB. In contrast, when offloading computations to the cloud, memory utilization remains nearly constant, indicating that AeroDaaS efficiently shifts the resource burden away from the edge when required. Despite these variations, the overall memory overhead remains modest ($\leq0.5$GB),i.e, only $8\%$ of the available memory, reinforcing AeroDaaS's efficiency and adaptability for real-time applications.

\subsubsection{\hilite{AeroDaaS eases the use of edge and cloud resources, and intelligent task schedulers for analytics, without additional overheads}}

\begin{figure}[!t]
    \centering
    \includegraphics[width=0.95\linewidth]{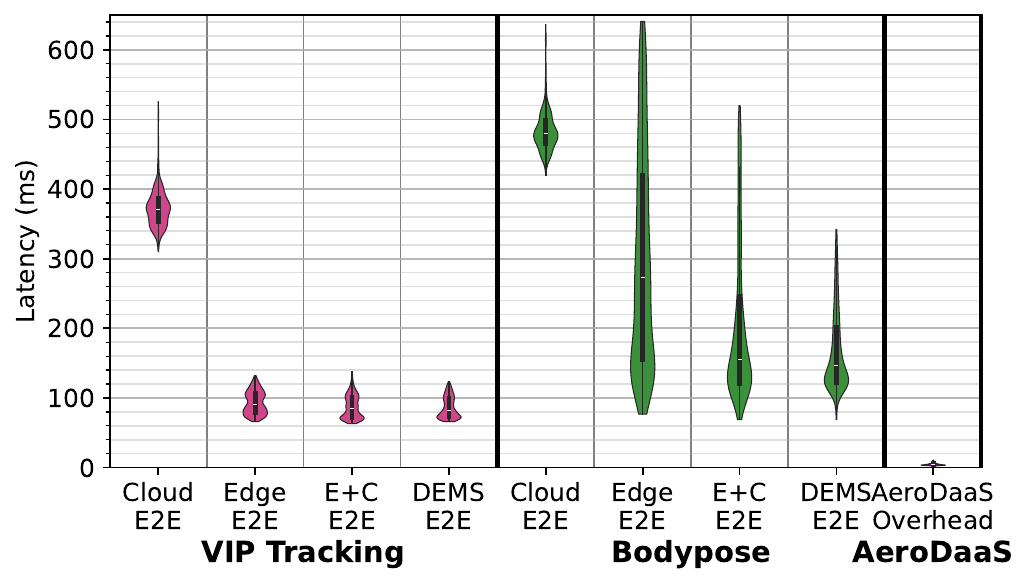}
    \vspace{-0.07in}
    \caption{AeroDaaS overhead and task scheduler latency using various scheduling algorithms for Situation Awareness application.
    }
    \label{fig:vip-nav-with-bodypose}
\end{figure}

\begin{figure}[!t]
    \centering
    \includegraphics[width=0.8\columnwidth]{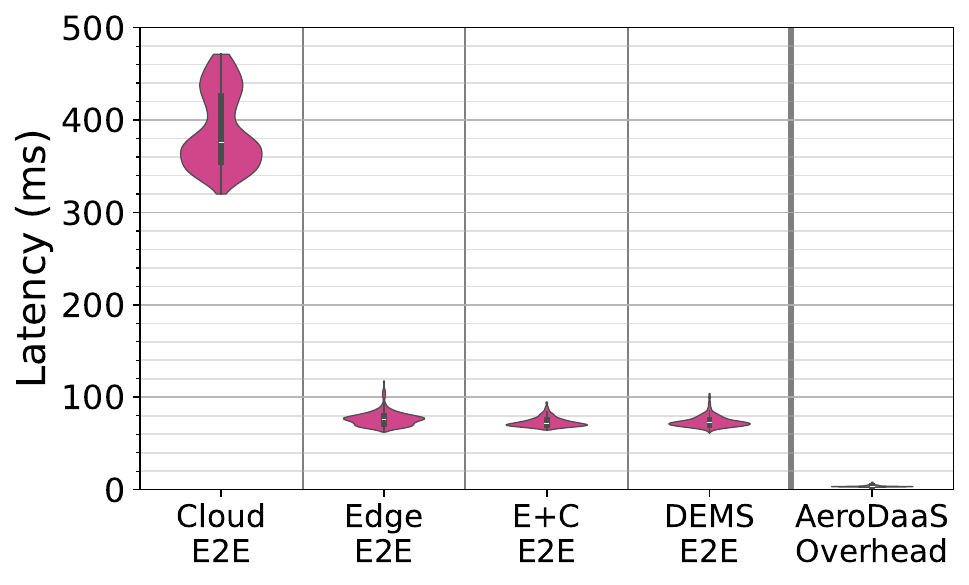}
    \caption{AeroDaaS overhead and task scheduler latency using various scheduling algorithms for VIP Tracking application.
    }
    \label{fig:vip-nav-without-bodypose}
\end{figure}

Beyond standalone edge and cloud execution, we integrate existing task scheduling techniques that enable dynamic edge-cloud offloading of DNN inference tasks~\cite{raj2024adaptiveheuristicsschedulingdnn}. Specifically, we incorporate edge-only, cloud-only, edge+cloud (E+C), and DEMS scheduling algorithms from this work. We evaluate the \textit{Situation Awareness} application using this scheduler, which involves YOLOv8 and a BodyPose estimation model. Fig.~\ref{fig:vip-nav-with-bodypose} presents the scheduler’s latency, accounting for both DNN inference and post-processing times. We observe significant variations in body pose estimation latency on the edge due to task queuing, a result of scheduling strategies. Notably, AeroDaaS maintains a consistent overhead, similar to cases without scheduling, reinforcing its flexibility and generalizability. Additionally, we evaluate the \textit{VIP Tracking} application under the same scheduling setup, reporting comparable findings in Fig.~\ref{fig:vip-nav-without-bodypose}.

\subsubsection{\hilite{AeroDaaS APIs allow applications to be concisely composed, within $40$ lines of code}}

\begin{figure}[!t]
    \centering
    \includegraphics[width=0.75\columnwidth]{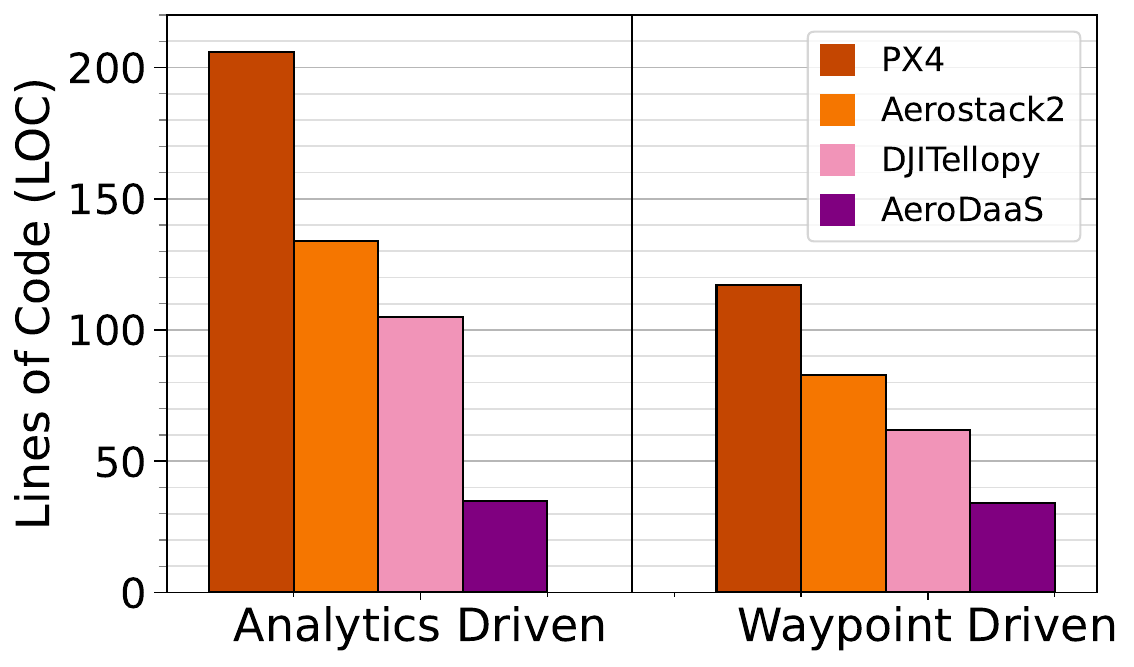}
    \caption{Lines of Code (LoC) required to implement analytics and waypoint driven applications using different frameworks.
    }
    \label{fig:loc-comparison}
    \vspace{-0.15in}
\end{figure}

AeroDaaS significantly reduces the Lines of Code (LoC), by upto $5\times$, required for users to compose UAV applications compared to other platforms, making it highly efficient and user-friendly. As shown in Fig.~\ref{fig:loc-comparison}, PX4 requires around $200$ LoC for analytics-driven applications (VIP Tracking) and over $130$ LoC for waypoint-driven applications (Farm Surveying), making it the most code-intensive. Aerostack2 reduces this overhead but still requires over $140$ LoC for analytics-driven tasks and more than $100$ LoC for waypoint-driven applications. Similarly, native DJI TelloPy code demands over $100$ LoC and around $70$ LoC for the two kind of tasks, respectively. In contrast, AeroDaaS requires fewer than $40$ LoC for both types of applications, offering the lowest complexity while maintaining flexibility and hardware-agnostic deployment. This is achieved by building on top of Aerostack2 and further abstracting away the APIs exposed to users, allowing seamless integration across different UAVs and computing platforms. With AeroDaaS, users can effortlessly compose and deploy applications while maintaining a minimal codebase. This makes it the most accessible choice, particularly for new users who want to quickly implement UAV applications with minimal effort.

\subsubsection{\hilite{Extensible analytics in AeroDaaS such as MonitoringAnalytics can help with easy visualization}}

\begin{figure}[!t]
\vspace{-0.1in}
    \centering
    \subfloat[Battery]{%
    \includegraphics[width=0.48\linewidth]{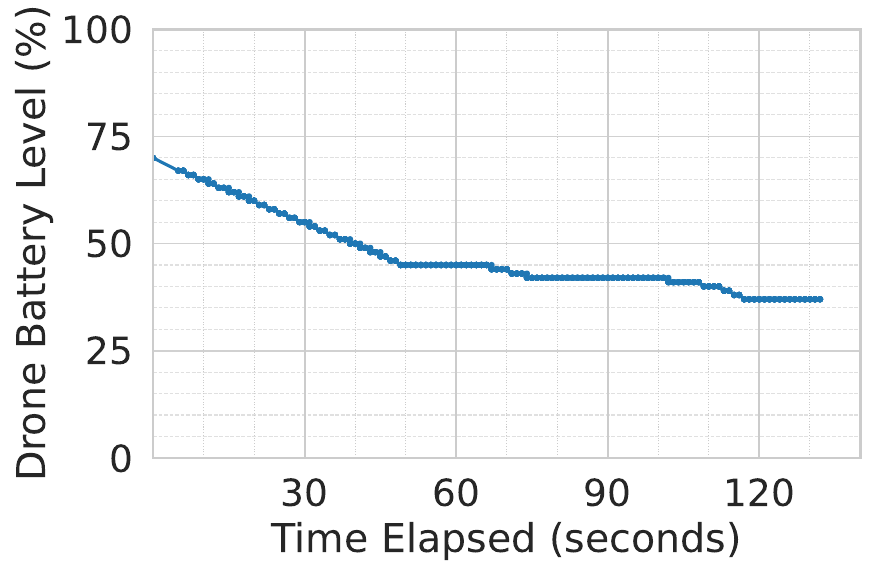}%
    \label{fig:battery-monitoring}%
    }~
    \subfloat[Altitude]{%
    \includegraphics[width=0.5\linewidth]{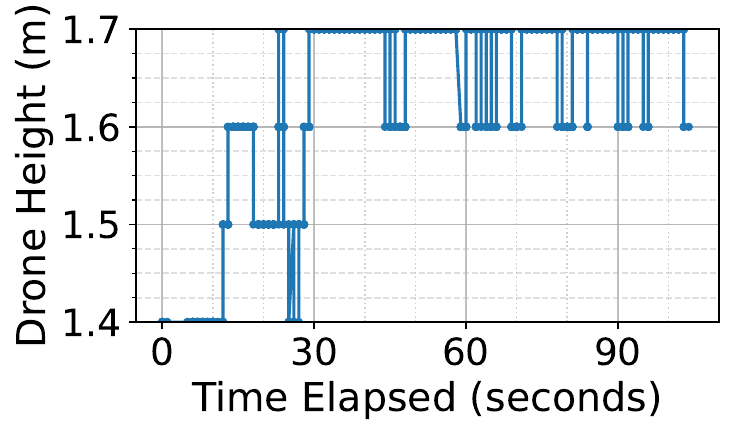}%
    \label{fig:height-monitoring}%
    }
\vspace{-0.05in}
\caption{Drone odometry data exposed using MonitoringAnalytics service for hardware experiments}
\label{fig:imonitoring-service}
\vspace{-0.15in}
\end{figure}

\begin{figure}[!t]
  \centering
\subfloat[Simulated crop survey application using PX4 SITL in Gazebo world]{
    \includegraphics[width=0.5\columnwidth]{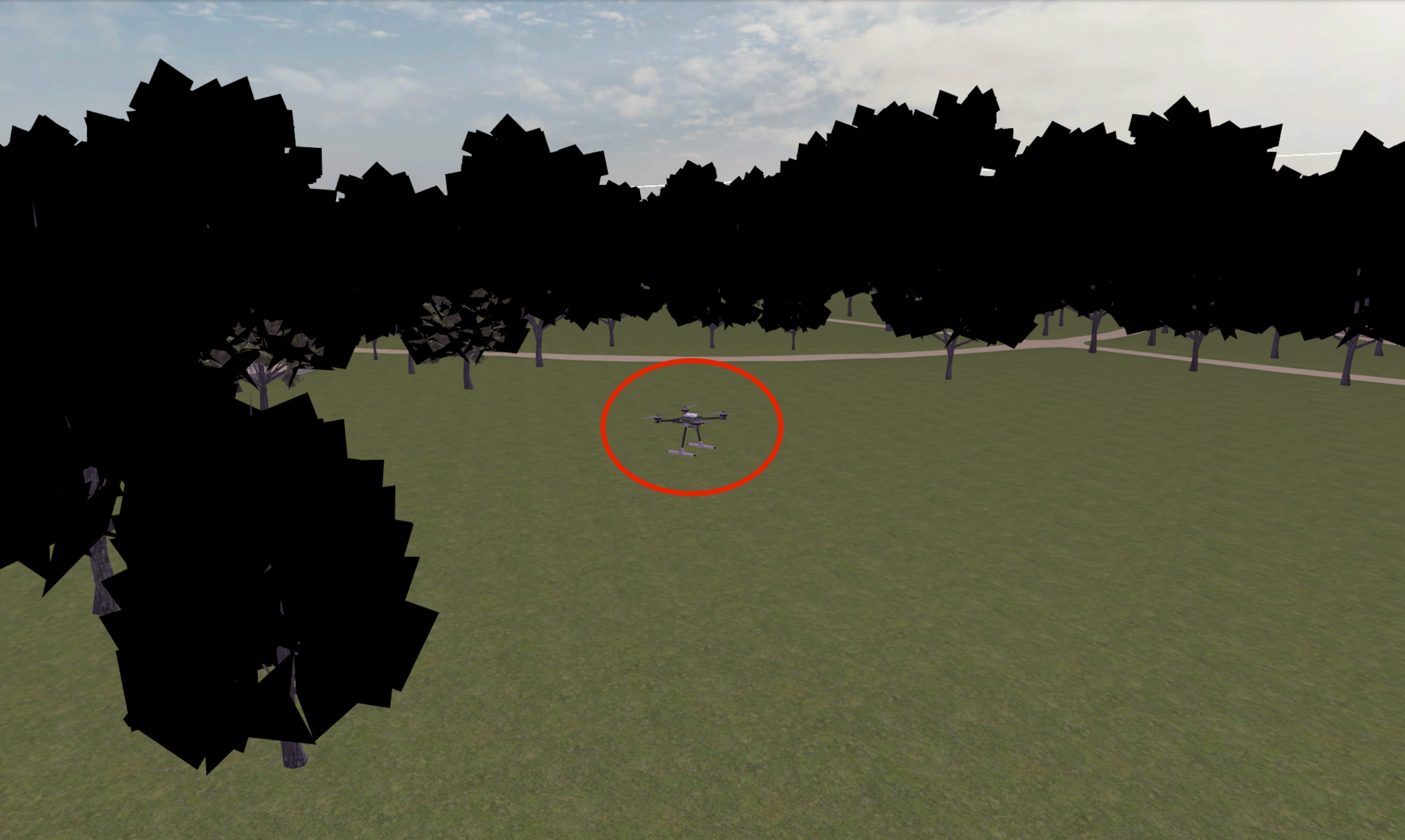}
   \label{fig:simulation_exp_survey}
  } \quad
  \subfloat[Trajectory data of the simulated drone]{
    \includegraphics[width=0.4\columnwidth]{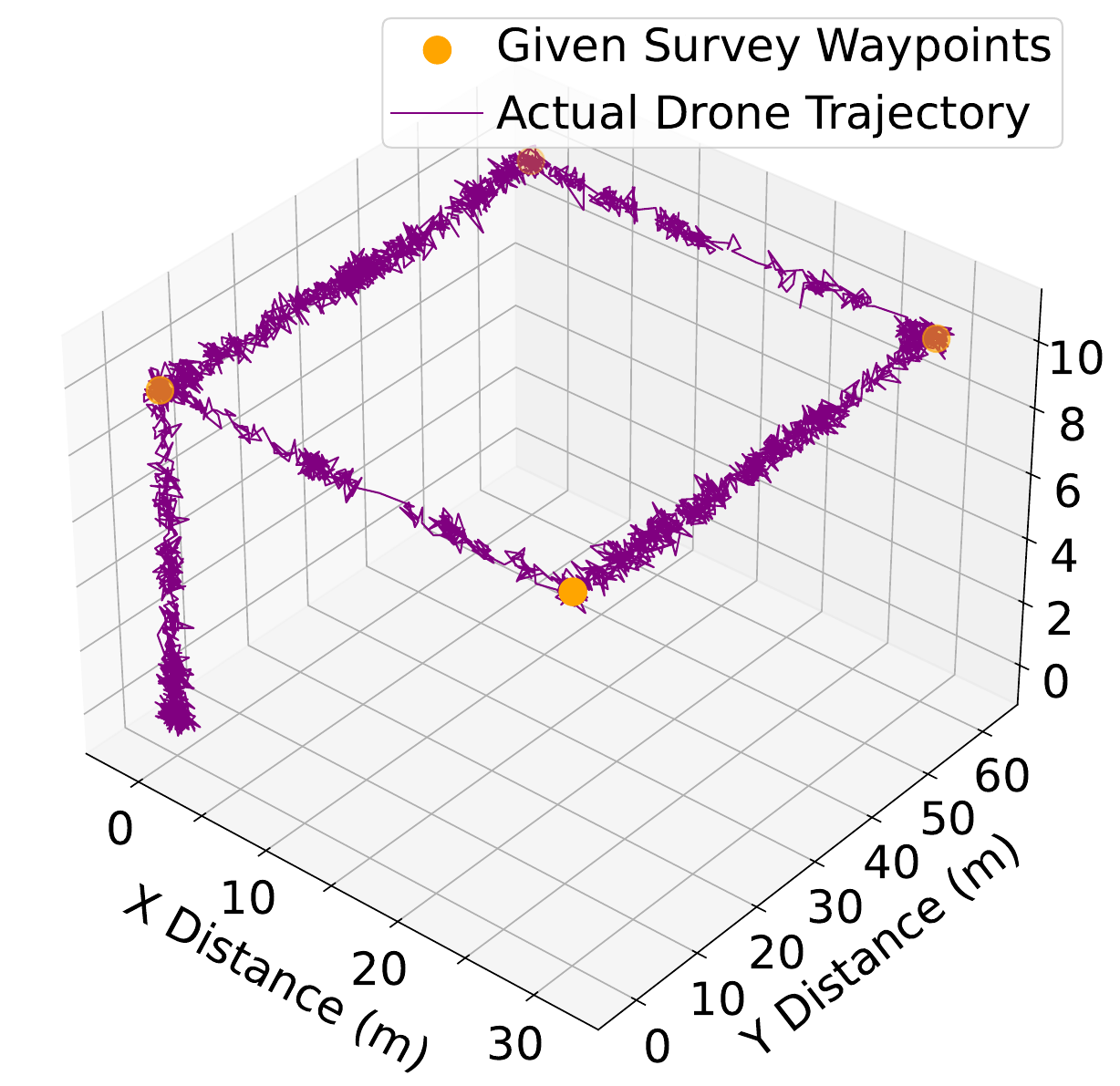}
   \label{fig:distance_simulation}}
	\caption{AeroDaaS simulation experiments    }
    \label{fig:AeroDaaS_evaluation_simulation}
    \vspace{-0.2in}
\end{figure}

We highlight the benefits of the \texttt{MonitoringAnalytics} service, which provides access to odometry and telemetry data that can be leveraged for downstream applications. For instance, Fig.\ref{fig:battery-monitoring} illustrates the consumption of the Tello drone’s battery over time as it follows the VIP in real-world scenarios. This information can be integrated into energy-aware drone handoff algorithms to optimize mission continuity. Similarly, Fig.\ref{fig:height-monitoring} presents the altitude variations of the Tello drone. Since the target height is set to $1.5$m for VIP tracking, we observe that the altitude oscillates between $1.4$m and $1.7$m. This real-time data can be fed into the PID controller to dynamically correct errors, ensuring robust tracking performance.

Additionally, we demonstrate the capabilities of \texttt{Monitoring\-Analytics} service in a simulated \textit{Farm Surveying} application simulated using Gazebo. The X500 drone (highlighted in red in Fig.\ref{fig:simulation_exp_survey}) surveys a rectangular area of $60m \times 30m$ at an altitude of $10$m. By exposing odometry readings, we visualize the drone’s trajectory in Fig.\ref{fig:distance_simulation}, where it takes off from the origin (0, 0, 0), follows a predefined set of waypoints (highlighted in orange), and returns to its starting position upon completing the survey.
It performs the 1 complete survey of the farm while running a pest detection DNN model on the input of the camera sensor and saving the output locally that can be accessed by the user later. This visualization helps users analyze the drone’s movement accuracy, assess waypoint adherence, and evaluate the efficiency of the surveying mission. 

\section{Related Work}\label{sec:related}

\begin{table*}[ht]
\centering
\caption{Comparison of Service-Oriented Frameworks for Drones-as-a-Service (DaaS) Applications}
\label{tab:related-work}
\footnotesize
\setlength{\tabcolsep}{2.5pt}
\begin{tabular}{l||c|c|c|c|c|c|c|c}
\toprule
\textbf{Framework} & \textbf{\makecell{Program.\\ Interface}} & \textbf{\makecell{Cross-\\Hard.}} & \textbf{\makecell{Navig-\\ation}} & \textbf{Sensing} & \textbf{Analytics} & \textbf{\makecell{Generali-\\zability}} & \textbf{\makecell{Edge +\\ Cloud}}  \\ \midrule
\textbf{UAL}   \cite{real2020unmanned}      & Python          & \checkmark & \checkmark & \checkmark & \texttimes & \checkmark & \texttimes  \\
\textbf{Aerostack2} \cite{fernandez2023aerostack2}  & Python          & \checkmark & \checkmark & \checkmark & \texttimes & \checkmark & \texttimes  \\
\textbf{Soft.Pilot} \cite{10.1145/3565386.3565484}  & Java          & \texttimes & \checkmark & \checkmark & \checkmark & \texttimes & \texttimes  \\
\textbf{SkyQuery} \cite{10.1145/3486607.3486750}    & Custom DSL      & \checkmark & \texttimes & \texttimes & \checkmark & \texttimes & \texttimes  \\
\textbf{Buzz} \cite{10.1109/IROS.2016.7759558}        & Buzz language   & \texttimes & \checkmark & \checkmark & \texttimes & \checkmark & \texttimes  \\
\textbf{VOLTRON} \cite{10.1145/2668332.2668353}    & Java and C++    & \checkmark & \texttimes & \checkmark & \texttimes & \checkmark & \texttimes  \\
\textbf{PaROS} \cite{10.1145/3197768.3197772}     & Java            & \checkmark & \checkmark & \checkmark & \texttimes & \checkmark & \texttimes  \\
\textbf{BeeCluster} \cite{10.1145/3386901.3388912} & Python          & \texttimes & \texttimes & \checkmark & \texttimes & \checkmark & \texttimes  \\
\textbf{AnDrone} \cite{van2019androne}    & Android Things  & \texttimes & \checkmark & \checkmark & \texttimes & \checkmark & \texttimes  \\ \midrule
\rowcolor[HTML]{DFFFD6}
\textbf{AeroDaaS}    & Python   & \checkmark & \checkmark & \checkmark & \checkmark & \checkmark & \checkmark \\ \bottomrule
\end{tabular}
\vspace{-0.1in}
\end{table*}

This section reviews existing frameworks, tools, and methodologies for UAV services, analyzing their contributions and limitations. We highlight how AeroDaaS fills key gaps by offering integrated, service-driven solutions for drone applications. Table~\ref{tab:related-work} summarizes the state-of-the-art service-oriented frameworks for UAVs.

\subsection{Cross-Platform UAV Programming}
Several frameworks provide low-level hardware abstractions for drones, focusing on control and actuation rather than higher-level services. AeroStack2~\cite{fernandez2023aerostack2} offers a ROS2-based abstraction layer, managing UAV hardware through a plugin-oriented architecture.  AeroDaaS builds on AeroStack2, automating infrastructure generation while extending capabilities to analytics integration, orchestration, and multi-device collaboration. UAV Abstraction Layer (UAL)~\cite{real2020unmanned} defines basic UAV functions but lacks extensibility and cross-platform compatibility. AeroDaaS provides a higher-level service layer, enabling analytics-driven applications and distributed coordination beyond UAL’s foundational controls. SoftwarePilot 2.0~\cite{10.1145/3565386.3565484} employs a microservices-based approach, containerizing mission logic using Docker and Kubernetes. However, its complex deployment and DJI-only support limit usability. AeroDaaS simplifies deployment while ensuring scalability, interoperability, and service-driven orchestration across diverse drone platforms.

\subsection{Domain Specific Languages (DSL) for UAVs}
Several frameworks leverage Domain-Specific Languages (DSLs) to abstract low-level drone operations. SkyQuery\cite{10.1145/3486607.3486750} enables video analytics and waypoint generation but lacks direct drone control services. Buzz\cite{10.1109/IROS.2016.7759558} provides a swarm programming DSL, but its reliance on the Buzz Virtual Machine (BVM) and absence of built-in analytics primitives add complexity. BeeCluster~\cite{10.1145/3386901.3388912} optimizes multi-drone missions using predictive analytics, yet lacks cross-platform compatibility.
AeroDaaS goes beyond DSL limitations by providing a service-driven, platform-agnostic framework that integrates analytics, navigation, and multi-device orchestration, enabling seamless UAV application development without low-level complexities.

\subsection{Programming Framework for UAVs}
VOLTRON\cite{10.1145/2668332.2668353} simplifies multi-drone coordination but is restricted to sensing services. PaROS\cite{10.1145/3197768.3197772} offers swarm orchestration primitives, automating specific missions but lacking DNN-based analytics services that leverage edge accelerators. AnDrone~\cite{van2019androne} virtualizes drones using Linux containers called Android Things. Users can configure virtual drones in the cloud, utilizing existing Android apps and resources, which are then safely deployed on real drone hardware. AnDrone is limited to Android apps, and offers just navigation and sensing capabilities. We use a similar Docker-based approach that can work across the edge and cloud continuum. 

\section{Conclusions and Future Work}
\label{sec:conclusions}
In this paper, we present a preliminary design for AeroDaaS, a programming framework to enable intuitive design, orchestration and management of drone-as-a-service application across the edge-cloud continuum. It abstracts low-level drone navigation controls, sensors and communication through high-level APIs to design drone applications across diverse drone hardware. It also integrates these with advanced user-defined analytics that can execute on edge and cloud resources. Early validation of AeroDaaS in done on two complex real-world DaaS applications executed using its APIs and runtime, both on real and simulated drones. These confirm that AeroDaaS eases development with minimal runtime overheads.

As future work, we plan to expand the AeroDaaS APIs to more fully support diverse application models and analytics/navigation patterns. We will focus on further enhancing the scalability, resilience and performance of AeroDaaS runtime while enabling multiple concurrent applications on the same drone with coordinated execution. We also aim to introduce primitives for fleet-level operations and evaluate AeroDaaS on additional drone platforms and complex applications.

\section*{Acknowledgements}
The first author was supported by a Prime Minister's Research Fellowship (PMRF) from the Government of India.

\balance

\bibliographystyle{IEEEtran} 
\bibliography{paper}

\end{document}